\documentclass[aps,prc,showpacs,superscriptaddress,twocolumn]{revtex4}

\usepackage{graphicx}

\begin{document}

\title{Ideal hydrodynamics and elliptic flow at SPS energies: Importance of the initial conditions}

\author{Hannah Petersen}
\affiliation{Frankfurt Institute for Advanced Studies (FIAS),
Ruth-Moufang-Str.~1, D-60438 Frankfurt am Main,
Germany}
\affiliation{Institut f\"ur Theoretische Physik, Johann Wolfgang Goethe-Universit\"at, Max-von-Laue-Str.~1, 
D-60438 Frankfurt am Main, Germany}

\author{Marcus Bleicher}
\affiliation{Institut f\"ur Theoretische Physik, Johann Wolfgang Goethe-Universit\"at, Max-von-Laue-Str.~1,
 D-60438 Frankfurt am Main, Germany}



\begin{abstract}
The elliptic flow excitation function calculated in a full (3+1)d hybrid Boltzmann approach with an intermediate hydrodynamic stage for heavy ion reactions from GSI-SIS to the highest CERN-SPS energies is discussed in the context of the experimental data. In this study, we employ a hadron gas equation of state to investigate the differences in the dynamics and viscosity effects. The specific event-by-event setup with initial conditions and freeze-out from a non-equilibrium transport model allows for a direct comparison between ideal fluid dynamics and transport simulations. At higher SPS energies, where the pure transport calculation cannot account for the high elliptic flow values, the smaller mean free path in the hydrodynamic evolution leads to higher elliptic flow values. In contrast to previous studies within pure hydrodynamics, the more realistic initial conditions employed here and the inclusion of a sequential final state hadronic decoupling provides results that are in line with the experimental data almost over the whole energy range from $E_{\rm lab}=2-160A~$GeV. Thus, this new approach leads to a substantially different shape of the $v_2/\epsilon$ scaling curve as a function of $(1/S dN_{ch}/dy)$ in line with the experimental data compared to previous ideal hydrodynamic calculations. This hints to a strong influence of the initial conditions for the hydrodynamic evolution on the finally observed $v_2$ values, thus questioning the standard interpretation that the hydrodynamic limit is only reached at RHIC energies. 
\end{abstract}

\keywords{Relativistic Heavy-ion collisions, Monte Carlo simulations, Hydrodynamic models, Collective flow}

\pacs{25.75.-q,24.10.Lx,24.10.Nz,25.75.Ld}

\maketitle

Transverse collective flow is one of the earliest predicted observables to probe heated and 
compressed nuclear matter \cite{Stoecker:1986ci,Voloshin:2008dg}. Elliptic flow, the anisotropy parameter that quantifies the momentum space anisotropy in the transverse plane of the outgoing particles of a heavy ion reaction, is a result of the pressure gradients that are present in the course of the evolution. Since it is a self-quenching effect, it is very sensitive to the early stage of the collision, i. e. the initial conditions and the mean free path during the high energy density stage of the evolution. 

Hydrodynamics has been proposed many years ago as a tool to describe collective effects in the hot and dense stage of heavy ion reactions where the matter might behave like a locally thermalized ideal fluid \cite{Rischke:1995ir,Rischke:1995pe,Csernai:1999nf,Hirano:2001eu,Kolb:2003dz,Nonaka:2006yn}. The hydrodynamic description has gained importance over the last few years because the high elliptic flow values that have been observed at the Relativistic Heavy Ion Collider (RHIC) seem compatible with some ideal hydrodynamic predictions \cite{Kolb:2000fha,Huovinen:2001cy}. At bombarding energies below the highest RHIC energies however, the same hydrodynamic calculations do overpredict the elliptic flow values by a large amount. This success of the hydrodynamic model has lead to the speculation that at RHIC energies and beyond the system reaches thermal equilibrium so quickly that a dense partonic (and nearly perfect) liquid could form. In this paper we investigate how robust this interpretation is, if proper initial conditions and freeze-out procedure are taken into account. This is necessary because hydrodynamic results depend strongly on the initial and final boundary conditions that are applied in the calculation. 

Previous calculations of elliptic flow in hadronic transport approaches have led to the conclusion that the pressure in the early stage of the collision is too low to reproduce the high elliptic flow values measured at RHIC \cite{Burau:2004ev,Petersen:2006vm,Lin:2001zk}. To get a more consistent picture of the whole dynamics of heavy ion reactions various so called microscopic plus macroscopic hybrid approaches have been launched during the last decade \cite{Nonaka:2006yn,Dumitru:1999sf,Bass:2000ib,Teaney:2000cw,Teaney:2001av,Grassi:2005pm,Andrade:2005tx,Hirano:2005xf,Hirano:2007ei,Andrade:2008xh,Andrade:2008fa}). Here we use the same technique and employ a transport approach with an embedded three-dimensional ideal relativistic one fluid evolution for the hot and dense stage of the reaction based on the Ultra-relativistic Quantum Molecular Dynamics (UrQMD) model \cite{Steinheimer:2007iy,Petersen:2008dd}. This approach allows to reduce the parameters for the initial conditions and provides a consistent freeze-out description and allows to compare the different underlying dynamics - ideal fluid dynamics vs. non-equilibrium transport - directly. 


This integrated Boltzmann+hydrodynamics transport approach is applied
 in this paper to simulate the dynamics of heavy ion collisions and to extract elliptic flow values from $E_{\rm lab}=2-160A~$GeV. To mimic experimental conditions as realistically as possible the initial
conditions and the final hadronic freeze-out are calculated using
the UrQMD approach. With this ansatz, the non-equilibrium dynamics in the
very early stage of the collision as well as the final state interactions
are properly taken into account on an event-by-event-basis.

Let us shortly describe the features of the present approach. 
UrQMD \cite{Bass:1998ca,Bleicher:1999xi,Petersen:2008kb} is a hadronic transport approach which simulates multiple
interactions of ingoing and newly produced particles, the excitation
and fragmentation of color strings and
the formation and decay of hadronic resonances. The coupling between the UrQMD initial state and the hydrodynamical
evolution proceeds when the two Lorentz-contracted nuclei have
passed through each other \cite{Steinheimer:2007iy}. Here, the spectators continue to propagate in the cascade and all other hadrons are mapped to the hydrodynamic grid. This treatment is especially important for non-central collisions which are of interest here. Event-by-event fluctuations are directly taken into account via event wise initial conditions generated by the primary collisions and string fragmentations in the microscopic UrQMD model. This leads to non-trivial velocity and energy density distributions for the hydrodynamical initial conditions as will be discussed below.

Starting from these initial conditions a full (3+1)
dimensional ideal hydrodynamic evolution is performed using the
SHASTA algorithm \cite{Rischke:1995ir,Rischke:1995mt}. The
hydrodynamic evolution is stopped, if the energy density
$\varepsilon$ drops below five times
the ground state energy density $\varepsilon_0$ (i.e. $\sim 730 {\rm
MeV/fm}^3$) in all cells. This criterion corresponds to a T-$\mu_B$-configuration
where the phase transition is expected (approximately $T=170$ MeV
at $\mu_B=0$). The hydrodynamic fields are then transformed to particle
degrees of freedom via the Cooper-Frye equation on an isochronous time $t$- hypersurface in the computational frame. The created particles proceed in their evolution in the hadronic cascade (UrQMD)
where rescatterings and final decays are calculated until all interactions cease and the system decouples. Further we refer to this kind of freeze-out
procedure as the isochronous freeze-out (IF).

Alternatively an approximate iso-eigentime freeze-out is chosen (see \cite{Li:2008qm} for details). Here, we freeze out full transverse
slices, of thickness $\Delta z = 0.2 $fm, whenever all cells in each individual slice fulfill the freeze-out criterion. For each slice we apply the
isochronous procedure described above. By doing
this one obtains a rapidity independent freeze-out temperature without artificial time dilatation effects. In
the following we will refer to this procedure as ``gradual
freeze-out''(GF). A more detailed description of the hybrid model
including parameter tests and results for multiplicities and spectra
can be found in \cite{Petersen:2008dd}.

Serving as an input for the hydrodynamical calculation the equation of state (EoS)
strongly influences the dynamics of an expanding system. In this
work we employ a hadron gas (HG) equation of state,
describing a non-interacting gas of free hadrons
\cite{Zschiesche:2002zr}. Included here are all reliably known
hadrons with masses up to $\approx 2$ GeV, which is equivalent to the active
degrees of freedom of the UrQMD model (note that this EoS does not
contain any form of phase transition). The purely hadronic
calculation serves as a baseline calculation to explore the effects
of the change in the underlying dynamics - pure transport vs.
hydrodynamic calculation.

\begin{figure}[ht]
\resizebox{0.6\textwidth}{!}{
\centering  
\includegraphics{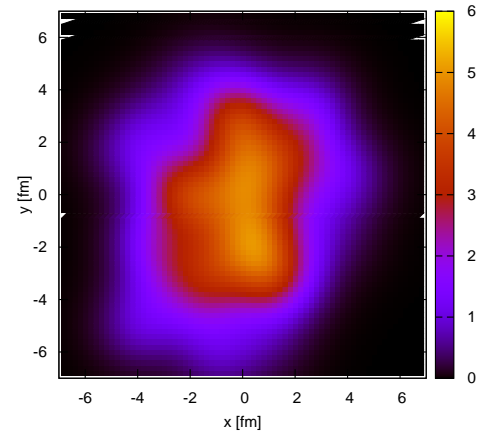}
}
\caption{(Color online) Energy density distribution in the $x-y$-plane for one midcentral ($b=7$ fm) Pb+Pb collision at $E_{\rm lab}=40A~$GeV.}
\label{edens}      
\end{figure}

\begin{figure}[ht]
\resizebox{0.6\textwidth}{!}{
\centering  
\includegraphics{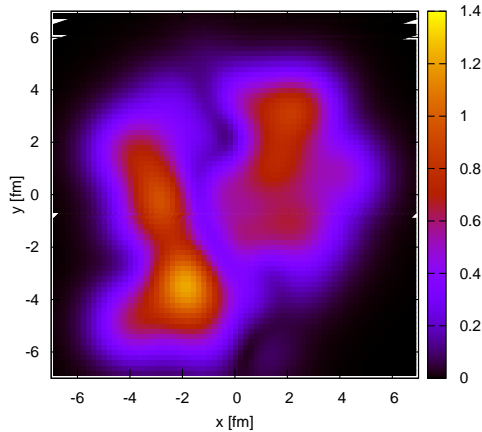}
}
\caption{(Color online) Velocity distribution in the transverse $x-y$-plane for one midcentral ($b=7$ fm) Pb+Pb collision at $E_{\rm lab}=40A~$GeV. The absolute value of the velocity has been multiplied by the energy density in the correspondig cell.}
\label{veldis}      
\end{figure}

We begin our investigation with the initial conditions for the hydrodynamical evolution. 
Fig. \ref{edens} shows the initial local rest frame energy density distribution in the transverse plane for one single Pb+Pb collision. The spatial anisotropy that causes the development of elliptic flow is nicely observed. The distribution is not smooth and not symmetric in any direction. In Fig. \ref{veldis} the initial velocity distribution is shown in the transverse plane. The value that is plotted here is the absolute value of the three-velocity of the hydrodynamic cells ($v=|\vec{v}|$) times the local rest frame energy density in the respective cell. In this way, one gets rid of the numerical noise in the almost empty cells. The velocity distribution is also not symmetric and fluctuates from event to event. As expected the velocities are higher at the edges of the almond shaped overlap region in $x$ direction. In the middle of the system the matter is almost at rest. To see how these space-momentum correlations transform to observables we introduce the elliptic flow $v_2$.   

The second coefficient of the Fourier expansion of the azimuthal distribution of the emitted particles ($v_2$) is called elliptic flow \cite{Sorge:1998mk,Ollitrault:1992bk,Bleicher:2000sx}. $v_2$ is defined by

\begin{equation}
v_2 \equiv \langle \rm{cos}[2(\phi-\Phi_{RP})]\rangle=\left\langle\frac{{\it p_x^2-p_y^2}}{{\it p_x^2+p_y^2}}\right\rangle  \quad,
\end{equation}

where $\phi$ is the azimuthal angle of the particle, $\Phi_{RP}$ is the azimuthal angle of the reaction plane and $p_x$ and $p_y$ are the momenta of the particle in $x$- and $y$-direction respectively.

\begin{figure}
\resizebox{0.5\textwidth}{!}{
\centering  
\includegraphics{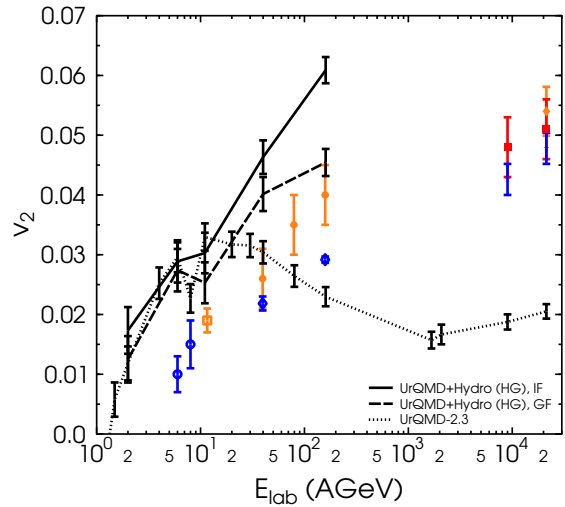}
}
\caption{(Color online) The energy excitation function of elliptic flow of charged particles in Au+Au/Pb+Pb 
collisions in mid-central collisions (b=5-9 fm) calculated at midrapidity ($|y|<0.5$) within the hybrid model with isochronous freeze-out (black full line) and gradual freeze-out (black dashed line) is contrasted to the pure UrQMD transport calculation (black dotted line). These curves are compared to data (colored symbols) from 
different experiments (E895, E877, NA49, CERES, PHENIX, PHOBOS and STAR) \cite{Alt:2003ab,v2data}. }
\label{flowexc}      
\end{figure}

In Fig. \ref{flowexc} we show the excitation function of charged particle elliptic flow compared to data over a wide energy range (Fig. \ref{flowexc}), i.e from $E_{\rm lab}=1A~$GeV to 
$\sqrt{s_{NN}}=200$~GeV. In this figure, the data and the calculation are not divided by further model dependent quantities and therefore it allows for a direct comparison. The symbols indicate the data for charged particles 
from different experiments. In the SPS regime the pure transport model calculations are quite in line with
the data, especially with the NA49 results. Above $E_{\rm lab}=160A~$GeV the calculation underestimates the elliptic 
flow. This has been taken as a sign that partonic degrees of freedom become more important at these energies. 

The smaller mean free path in the hybrid model calculation leads to higher elliptic flow values at higher SPS energies even without explicit phase transition. At lower energies the result is in line with the transport calculation since the hydrodynamic evolution is very short. The average duration of the hydrodynamic evolution increases from $\sim 3$ fm at low energies to around 8 fm at $E_{\rm lab}=40A~$GeV and even $\sim 12$ fm at the highest SPS energy. Please note that the crucial observation is not only that there is higher elliptic flow than in the transport calculation, but that the hybrid approach shows that ideal hydrodynamics is less than 20 \% away from the experimental data. This confirms that the initial conditions and the freeze-out treatment have important influence on the results of a hydrodynamic calculation. 

\begin{figure}
\resizebox{0.5\textwidth}{!}{
\centering  
\includegraphics{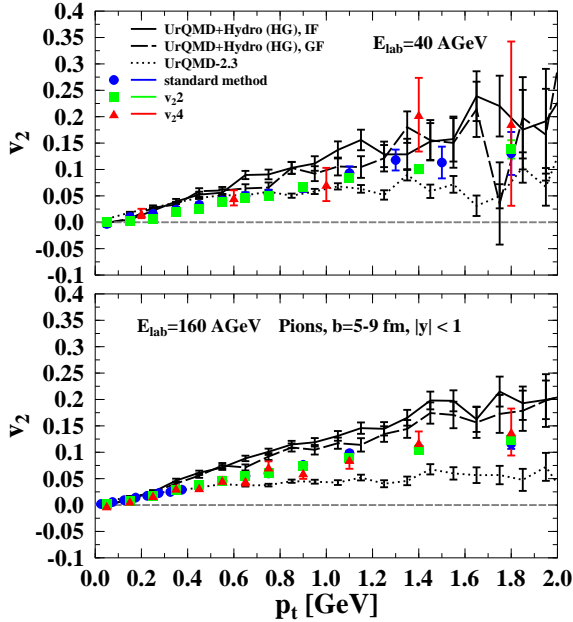}
}
\caption{(Color online) Elliptic flow of pions in mid-central (b=5-9) Pb+Pb collisions at $E_{\rm lab}=40A~$GeV and  $E_{\rm lab}=160A~$GeV. The full black line depicts the hybrid model calculation with isochronous freeze-out, the black dashed line the hybrid calculation employing the gradual freeze-out while the pure transport calculation is shown as the black dotted line. The colored symbols display experimental data obtained with different measurement methods by NA49 \cite{Alt:2003ab}. }
\label{v2_ptpi}      
\end{figure}

The elliptic flow of pions as a function of transverse momentum is shown in Fig. \ref{v2_ptpi}. As it has been stated above the hydrodynamic evolution leads to higher elliptic flow values especially at higher $p_t$ where the pure transport calculation underpredicts the data. For these differential results at midrapidity the difference between the two freeze-out prescriptions is less pronounced than for the integrated results. The hybrid model calculation leads also for the differential elliptic flow to a reasonable agreement with the experimental data.  

\begin{figure}
\resizebox{0.5\textwidth}{!}{
\centering  
\includegraphics{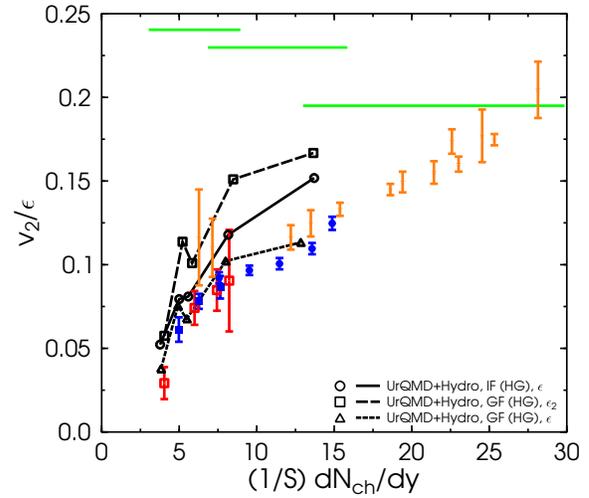}
}
\caption{(Color online) $v_2/\epsilon$ as a function of $(1/S)dN_{\rm ch}/dy$ for different energies and centralities for Pb+Pb/Au+Au collisions. The results from mid-central collisions (b=5-9 fm) calculated within the hybrid model with isochronous/gradual freeze-out (full line with circles and dashed line with triangles respectively) are shown. Furthermore, the hybrid model calculation with GF is divided by a different eccentricity ($\epsilon_{2}$)(dashed line with squares). These curves are compared to data depicted by colored symbols from  different experiments (E877, NA49 and STAR \cite{Alt:2003ab}) for mid-central collisions. The green full lines correspond to the previously calculated hydrodynamic limits \cite{Kolb:2000sd}.}
\label{scaling}      
\end{figure}

Finally, we replot the $v_2 (\sqrt{s}_{NN})$ values as a function of particle density and scaled by the eccentricity of the initial state. Fig. \ref{scaling} shows $v_2/\epsilon$ as a function of $(1/S) dN_{\rm ch}/dy$ which is assumed to be a decreasing quantity in the investigated energy regime in ideal hydrodynamics calculations. $(1/S) dN_{\rm ch}/dy$ is the charged particle density at midrapidity divided by the initial state overlap area. This way of plotting the elliptic flow excitation function allows to compare results from different energies and centralities at the same time. The charged particle multiplicity in the overlap area is the same in a central low energy collision as in a peripheral high energy collision. The calculations within the UrQMD + hydrodynamics approach have been performed for midcentral Pb+Pb/Au+Au collisions with $b=5-9$ fm. The charged particle multiplicities and $v_2$ values have been calculated at midrapidity ($|y| < 0.5$). For the evaluation of the relevant initial eccentricity and the overlap area we have stopped the calculation of 10.000 UrQMD events at the time of the overlap of the nuclei. The quantities of interest have been evaluated for the participating particles which are defined as the nucleons which have undergone at least one interaction plus the newly produced particles (the result is insensitive to the fact if only nucleons are considered or not) according to the following formula \cite{nucl-th/9906075,Sorge:1998mk}:
 
\begin{equation}
\epsilon = \frac{\langle y^2 \rangle-\langle x^2\rangle}{\langle y^2 \rangle+\langle x^2 \rangle} \quad \mbox{and } S=\pi \sqrt{\langle x^2 \rangle \langle y^2 \rangle}\quad,
\end{equation}  

where the averages are taken over particles and events at the same time. An alternative defintion is sometimes used \cite{Voloshin:2008dg}

\begin{equation}
\epsilon_{2}=\frac{\langle y^2-x^2\rangle}{\langle y^2+x^2 \rangle}
\end{equation} 

where the averages are taken first over particles in one event and then the value for $\epsilon$ is averaged over events. In this way, the events with higher particle production have the same weight as those with lower multiplicities. 

Since it is not obvious which way of calculating the initial eccentricity captures the physical picture best (e.g. the Glauber values which are taken from experiment) we show both possibilities because they lead to different results. 

It is remarkable that the shape of the curve is substantially changed in the hybrid model calculation compared to the previous calculated hydro limits \cite{Kolb:2000sd} (shown as horizontal lines). A similar shape has also been obtained in a two-dimensional hybrid calculation with a hadronic afterburner but simplified initial conditions and a EoS including a phase transition to the QGP (see \cite{Teaney:2000cw}). The present calculation with ideal one fluid hydrodynamics with hadronic degrees of freedom as described above is however able to reproduce the shape of the experimental data points and even the magnitude at lower energies. Note that at very low energies the hydrodynamic stage is rather short and does not influence the evolution considerably. In this regime all different setups show the same results and are compatible with the data. Towards higher energies (where pure hadronic transport calculations have too much viscosity and underpredict the data), the hybrid calculation leads to higher pressure gradients in the early stage and therefore to suitable elliptic flow values. Most important here is the only moderate increase in $v_2$ because of the more realistic treatment via the inclusion of initial non-equilibrium effects and a complex shape (both in coordinate and momentum space) of the initial energy and baryon density distribution in addition to our sophisticated final state freeze-out. In fact, for the (more physical) gradual freeze-out, the $v_2$ values are reduced by more than a factor of 2 compared to ideal hydrodynamics with simplified initial conditions. The influence of the freeze-out prescription can be observed by comparing the isochronous to the gradual freeze-out scenario. The alternative calculation of the eccentricity ($\epsilon_2$) leads to higher values because $\epsilon_2$ is in general smaller than $\epsilon$. Additional changes in magnitude may be caused by viscosity effects during the hot and dense stage and a possible phase transition which might weaken the pressure gradients in the early stage.

We have shown that the elliptic flow values at SPS energies are in line with an ideal hydrodynamic evolution if a proper initial state is used and the final freeze-out proceeds gradually. An integrated hybrid model calculation with initial conditions and freeze-out from microscopic transport with an embedded (3+1)d hydrodynamic evolution is able to reproduce the experimentally measured $v_2$ values. This points to the fact that the treatment of initial conditions and freeze-out is crucial for any hydrodynamic calculation.

\section{Acknowledgements}
\label{ack}
We are grateful to the Center for the Scientific Computing (CSC) at Frankfurt for the computing resources. The authors thank Dirk Rischke for providing the 1 fluid hydrodynamics code. The authors thank Gunnar Gr\"af, Raimond Snellings, Pasi Huovinen and Michael Mitrovski for fruitful discussions. H. Petersen gratefully acknowledges financial support by the Deutsche Telekom-Stiftung and support from the Helmholtz Research School on Quark Matter Studies. This work was supported by GSI and BMBF. This work was supported by the Hessian LOEWE initiative through the Helmholtz International Center for FAIR (HIC for FAIR).



\begin{thebibliography}{99}

\bibitem{Stoecker:1986ci}
  H.~Stoecker and W.~Greiner,
  Phys.\ Rept.\  {\bf 137}, 277 (1986).

\bibitem{Voloshin:2008dg}
  S.~A.~Voloshin, A.~M.~Poskanzer and R.~Snellings,
  arXiv:0809.2949 [nucl-ex].

\bibitem{Rischke:1995ir}
  D.~H.~Rischke, S.~Bernard and J.~A.~Maruhn,
  Nucl.\ Phys.\  A {\bf 595}, 346 (1995)

\bibitem{Rischke:1995pe}
  D.~H.~Rischke, Y.~Pursun, J.~A.~Maruhn, H.~Stoecker and W.~Greiner,
  Heavy Ion Phys.\  {\bf 1}, 309 (1995)

\bibitem{Csernai:1999nf}
  L.~P.~Csernai and D.~Rohrich,
  Phys.\ Lett.\  B {\bf 458}, 454 (1999)

\bibitem{Hirano:2001eu}
  T.~Hirano,
  Phys.\ Rev.\  C {\bf 65}, 011901 (2002)

\bibitem{Kolb:2003dz}
  P.~F.~Kolb and U.~W.~Heinz,
  arXiv:nucl-th/0305084.

\bibitem{Nonaka:2006yn}
  C.~Nonaka and S.~A.~Bass,
  Phys.\ Rev.\  C {\bf 75}, 014902 (2007)

\bibitem{Kolb:2000fha}
  P.~F.~Kolb, P.~Huovinen, U.~W.~Heinz and H.~Heiselberg,
  Phys.\ Lett.\  B {\bf 500}, 232 (2001)

\bibitem{Huovinen:2001cy}
  P.~Huovinen, P.~F.~Kolb, U.~W.~Heinz, P.~V.~Ruuskanen and S.~A.~Voloshin,
  Phys.\ Lett.\  B {\bf 503}, 58 (2001)

\bibitem{Burau:2004ev}
  G.~Burau, J.~Bleibel, C.~Fuchs, A.~Faessler, L.~V.~Bravina and E.~E.~Zabrodin,
  Phys.\ Rev.\  C {\bf 71}, 054905 (2005)

\bibitem{Petersen:2006vm}
  H.~Petersen, Q.~Li, X.~Zhu and M.~Bleicher,
  Phys.\ Rev.\  C {\bf 74}, 064908 (2006)

\bibitem{Lin:2001zk}
  Z.~w.~Lin and C.~M.~Ko,
  Phys.\ Rev.\  C {\bf 65}, 034904 (2002)

\bibitem{Dumitru:1999sf}
  A.~Dumitru, S.~A.~Bass, M.~Bleicher, H.~Stoecker and W.~Greiner,
  Phys.\ Lett.\  B {\bf 460}, 411 (1999)

\bibitem{Bass:2000ib}
  S.~A.~Bass and A.~Dumitru,
  Phys.\ Rev.\  C {\bf 61}, 064909 (2000)

\bibitem{Teaney:2000cw}
  D.~Teaney, J.~Lauret and E.~V.~Shuryak,
  Phys.\ Rev.\ Lett.\  {\bf 86} (2001) 4783


\bibitem{Teaney:2001av}
  D.~Teaney, J.~Lauret and E.~V.~Shuryak,
  arXiv:nucl-th/0110037.

\bibitem{Grassi:2005pm}
  F.~Grassi, Y.~Hama, O.~Socolowski and T.~Kodama,
  J.\ Phys.\ G {\bf 31}, S1041 (2005).

\bibitem{Andrade:2005tx}
  R.~Andrade, F.~Grassi, Y.~Hama, T.~Kodama, O.~.~J.~Socolowski and B.~Tavares,
  Eur.\ Phys.\ J.\  A {\bf 29}, 23 (2006)

\bibitem{Hirano:2005xf}
  T.~Hirano, U.~W.~Heinz, D.~Kharzeev, R.~Lacey and Y.~Nara,
  Phys.\ Lett.\  B {\bf 636}, 299 (2006)

\bibitem{Hirano:2007ei}
  T.~Hirano, U.~W.~Heinz, D.~Kharzeev, R.~Lacey and Y.~Nara,
  Phys.\ Rev.\  C {\bf 77}, 044909 (2008)

\bibitem{Andrade:2008xh}
  R.~P.~G.~Andrade, F.~Grassi, Y.~Hama, T.~Kodama and W.~L.~Qian,
  Phys.\ Rev.\ Lett.\  {\bf 101}, 112301 (2008)

\bibitem{Andrade:2008fa}
  R.~P.~G.~Andrade, A.~L.~V.~Reis, F.~Grassi, Y.~Hama, W.~L.~Qian, T.~Kodama and J.~Y.~Ollitrault,
  arXiv:0812.4143 [nucl-th].

\bibitem{Steinheimer:2007iy}
  J.~Steinheimer, M.~Bleicher, H.~Petersen, S.~Schramm, H.~Stocker and D.~Zschiesche,
  Phys.\ Rev.\  C {\bf 77}, 034901 (2008)

\bibitem{Petersen:2008dd}
  H.~Petersen, J.~Steinheimer, G.~Burau, M.~Bleicher and H.~Stocker,
  Phys.\ Rev.\  C {\bf 78}, 044901 (2008)

\bibitem{Bass:1998ca}
  S.~A.~Bass {\it et al.},
  Prog.\ Part.\ Nucl.\ Phys.\  {\bf 41}, 255 (1998)
  [Prog.\ Part.\ Nucl.\ Phys.\  {\bf 41}, 225 (1998)]

\bibitem{Bleicher:1999xi}
  M.~Bleicher {\it et al.},
  J.\ Phys.\ G {\bf 25}, 1859 (1999)

\bibitem{Petersen:2008kb}
  H.~Petersen, M.~Bleicher, S.~A.~Bass and H.~Stocker,
  arXiv:0805.0567 [hep-ph].

\bibitem{Rischke:1995mt}
  D.~H.~Rischke, Y.~Pursun and J.~A.~Maruhn,
  Nucl.\ Phys.\  A {\bf 595}, 383 (1995)
  [Erratum-ibid.\  A {\bf 596}, 717 (1996)]

\bibitem{Li:2008qm}
  Q.~Li, J.~Steinheimer, H.~Petersen, M.~Bleicher and H.~Stoecker,
  arXiv:0812.0375 [nucl-th].

\bibitem{Zschiesche:2002zr}
  D.~Zschiesche, S.~Schramm, J.~Schaffner-Bielich, H.~Stoecker and W.~Greiner,
  Phys.\ Lett.\  B {\bf 547}, 7 (2002)

\bibitem{Sorge:1998mk}
  H.~Sorge,
  Phys.\ Rev.\ Lett.\  {\bf 82}, 2048 (1999)

\bibitem{Ollitrault:1992bk}
  J.~Y.~Ollitrault,
  Phys.\ Rev.\  D {\bf 46}, 229 (1992).

\bibitem{Bleicher:2000sx}
  M.~Bleicher and H.~Stoecker,
  Phys.\ Lett.\  B {\bf 526}, 309 (2002)



\bibitem{Alt:2003ab}
  C.~Alt {\it et al.}  [NA49 Collaboration],
  Phys.\ Rev.\  C {\bf 68}, 034903 (2003)


\bibitem{v2data}
  C.~Pinkenburg {\it et al.}  [E895 Collaboration],
{\it Prepared for Centennial Celebration and Meeting of the American Physical Society (Combining Annual APS General Meeting and the Joint Meeting
of the APS and the AAPT), Atlanta, Georgia, 20-26 Mar 1999}

  P.~Chung {\it et al.}  [E895 Collaboration],
  Phys.\ Rev.\  C {\bf 66}, 021901 (2002)


  K.~Filimonov {\it et al.}  [CERES/NA45 Collaboration],
  arXiv:nucl-ex/0109017.

  J.~Slivova  [CERES/NA45 Collaboration],
  Nucl.\ Phys.\  A {\bf 715}, 615 (2003)


S.I.Esumi, J. Slivova, J. Milosevic for CERES Collaboration
SFIN, year XV, Series A: Conferences, No. A2(2002)

  S.~Esumi  [PHENIX Collaboration],
  Nucl.\ Phys.\  A {\bf 715}, 599 (2003)

  S.~Manly {\it et al.}  [PHOBOS Collaboration],
  Nucl.\ Phys.\  A {\bf 715}, 611 (2003)

  R.~L.~Ray  [STAR Collaboration],
  Nucl.\ Phys.\  A {\bf 715}, 45 (2003)

\bibitem{Kolb:2000sd}
  P.~F.~Kolb, J.~Sollfrank and U.~W.~Heinz,
  Phys.\ Rev.\  C {\bf 62}, 054909 (2000)

\bibitem{nucl-th/9906075}
  S.~A.~Voloshin and A.~M.~Poskanzer,
  Phys.\ Lett.\  B {\bf 474}, 27 (2000)







\end{thebibliography}
\end{document}